\documentclass[preprint,12pt]{elsarticle}

\usepackage{amssymb}
 \usepackage{amsthm,amssymb,amsfonts}
\usepackage{epsfig}
\usepackage{epstopdf}
\usepackage{subfigure}
\usepackage{array}
\usepackage{changes}
\usepackage{graphicx}
\usepackage{textcomp}
\usepackage{xcolor}
\usepackage{amsmath}
\usepackage{fancyhdr}
\usepackage{textcomp}
\usepackage{gensymb}
\usepackage{makecell}
\usepackage{algorithm}
\usepackage{algpseudocode}

\journal{Digital Signal Processing}

\begin{document}

\begin{frontmatter}



\title{On Smart Morphing Wing Aircraft Robust Adaptive Beamforming}


\author[]{Yizhen~Jia}
\ead{jiayizhen@std.uestc.edu.cn}
\author[]{Hui Chen}
\ead{huichen0929@uestc.edu.cn}
\author[]{Wen-Qin Wang\corref{mycorrespondingauthor}}
\cortext[mycorrespondingauthor]{Corresponding author}
\ead{ wqwang@uestc.edu.cn}
\author[]{Jie~Cheng}
\ead{chengjie@std.uestc.edu.cn}

\address{School of  Information and Communication Engineering, University of Electronic Science and Technology of China, Chengdu, $611731$, China}
\fntext[]{This work was supported by National Natural Science Foundation of
	China 62171092 and Sichuan Science and Technology Program under grant
	2018RZ0141.}
\begin{abstract}
	The smart morphing wing aircraft (SMWA) is a highly adaptable platform that can be widely used for intelligent warfare due to its real-time variable structure. The flexible conformal array (FCA) is a vital detection component of SMWA, when the deformation parameters of FCA are mismatched or array elements are mutually coupled, detection performance will be degraded. To overcome this problem and ensure robust beamforming for FCA, deviations in array control parameters (ACPs) and array perturbations, the effect of mutual coupling in addition to looking-direction errors should be considered. In this paper, we propose a robust adaptive beamforming (RAB) algorithm by reconstructing a multi-domain interference plus noise covariance matrix (INCM) and estimating steering vector (SV) for FCA. We first reconstruct the INCM using multi-domain processing, including ACP and angular domains. Then, SV estimation is executed through an optimization procedure. Experimental results have shown that the proposed beamformer outperforms existing beamformers in various mismatch conditions and harsh environments, such as high interference-to-noise ratios, and mutual coupling of antennas.  

\end{abstract}




\end{frontmatter}


\section{Introduction}
S{\scshape mart} morphing wing aircraft (SMWA) has attracted considerable interest from scholars in the aerospace vehicle community \cite{CHU2022220,Barbarino2011ARO}, due to the adaptive variable structural geometry based on the scenario and improved aerodynamic efficiency. Therefore, flexible conformal arrays (FCA), as an important part of SMWA, have received increasing attention from the research community in recent years \cite{HashemiMohammedRezaM,9338189,9956877,Jia9864349}. Due to its flexibility and conformal advantages, it allows a wide range of novel applications \cite{9330010,6338271,8701107,7119183}. For example, the physical surface of a conformal array is used to increase the effective antenna aperture, while reducing the array volume and weight. Additionally, the flexibility makes it easier for the FCA to generate the desired antenna beams for a variety of situations \cite{6338271}. Emerging applications require capabilities
beyond conventional static conformal arrays, such as using FCA in wearables and lightweight devices \cite{HashemiMohammedRezaM}, deployable apertures that dynamically change shape \cite{9338189}. Although FCAs can constantly change shape during deployment or operation, their element positions and mutual coupling may not be known in the practical working environment, which would result in significant performance degradation. Therefore, some robust adaptive beamforming (RAB) methods should be proposed to compensate for this degradation with respect to FCA.

As is well known, adaptive beamforming technology is very sensitive to various
mismatches, such as array calibration errors, incoherent local scattering, wavefront distortion and
direction-of-arrival (DOA) errors \cite{0471733482}, resulting in significant performance degradation in the output signal-to-interference-plus-noise ratio (SINR). In addition, the adaptive beamforming method for FCA must also be robust to the inaccuracy of array control parameters (ACPs) during deformable operations of FCA, which will trigger the steering vector (SV) mismatch. Clearly, this type of error is unique to FCA, which has not been fully investigated in recent literature. Although ACP errors can be reduced to SV errors, the law of the error has not been fully discussed and the effectiveness of conventional RAB methods applied to this situation has not been investigated. Over the past decades, substantial RAB approaches have been studied \cite{YANG2023103977,8642908,9230850,9545772,650188,5510181,4610274,6180020,8642927,9358208,7446358,1166614,Yang9246694} to mitigate the effects of model mismatches and improve beamformers' robustness. However, the performance of these methods suffers significant performance degradation when they are applied directly to FCAs. For FCAs, the ACP errors induce severe SV mismatch, and additional looking direction error, as well as the ill  interference-plus-noise covariance matrix (INCM) due to the echo signal received by time-varying geometry, which motivates us to address this thorny issue. 

In general, the proposed RAB techniques can be broadly classified into four categories---diagonal loading (DL)-based (\cite{8642908,9230850}), eigenspace-based (\cite{9545772,650188}), uncertainty set constraints-based (\cite{5510181,4610274}), INCM reconstruction and SV estimation-based approaches (\cite{YANG2023103977,6180020,8642927,Yang9246694,Peng9534656}), and some other techniques (\cite{9358208,7446358,1166614}). Inspired by \cite{6180020,1166614} and our previous work \cite{Jia9864349}, we extend the INCM-based framework to address the FCA RAB problem. Specifically, we propose a RAB approach based on multiple domain INCM reconstruction and SV estimation in this paper to reduce the impact of ACP errors along with other types of errors (i.e., looking direction error and array perturbation).
INCM is reconstructed not only on the angle domain, but also on the ACP domain, which is the extension of the method in \cite{6180020}. Then, an alternative optimization scheme is proposed to estimate the true SV in the presence of ACP errors and array perturbation with mutual coupling. 
In short, we first reconstruct INCM with discrete summation of rank-one matrix in angle and ACP domains, and then estimate SV by solving two alternative optimization problems. 
The true SV is composed of two terms, one term is the orthogonal component that can be estimated by a QCQP problem, and the other term is the perturbation component that can be estimated by solving a barrier function using the internal point method.

The rest of the paper is organized as follows: Section II introduces the FCA geometry and signal model with mutual coupling effects, and summarizes the FCA optimization object in case of mixed mismatches in practice. Next, the RAB algorithm is proposed in Section III. Numerical simulations of three scenarios are provided in Section IV and conclusions are drawn in Section V.

\textbf{\emph{notations:}} $ \odot  $ represents the Hadamard Product, $ \mathrm{diag}\left\{  \cdot  \right\}  $ denotes the diagonalize matrix of brace. $ \left( \bullet \right) ^T $ and $ \left( \bullet \right) ^H $ denote the transpose and Hermitian transpose, respectively. $ E\left\{ \bullet \right\}  $ is the statistical expectation. $ | \bullet|$ is the absolute value, and $ \left\| \bullet \right\|_2 $ is the norm-2 of the vector or matrix.

\section{Signal Model of FCA}
\label{Se2}

\begin{figure}[htp]
	\centering
	\subfigure {\includegraphics[width=0.7\textwidth]{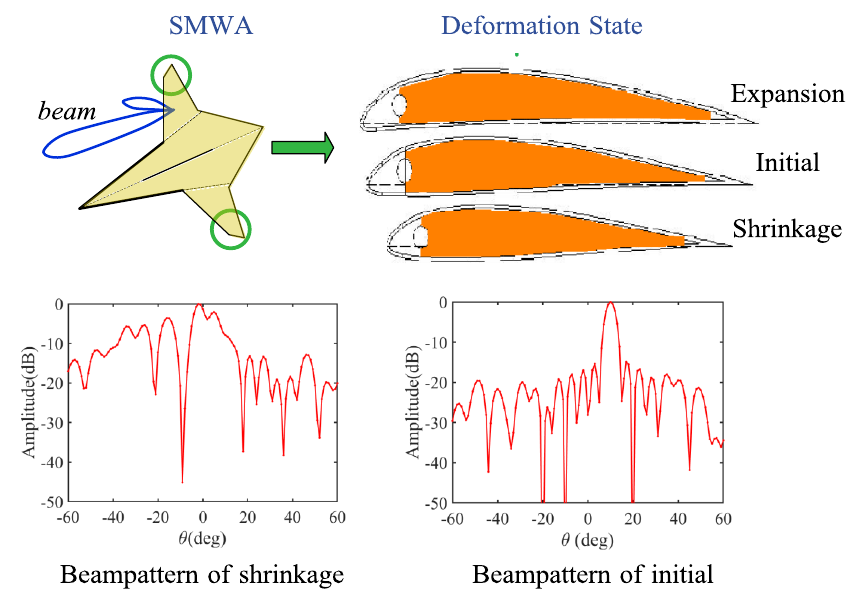}}
	\caption{ The deformation scenarios of SMWA and corresponding beampatterns.}
	\label{illustrate}
\end{figure}

\begin{figure}[htp]
	\centering
	\subfigure {\includegraphics[width=0.6\textwidth]{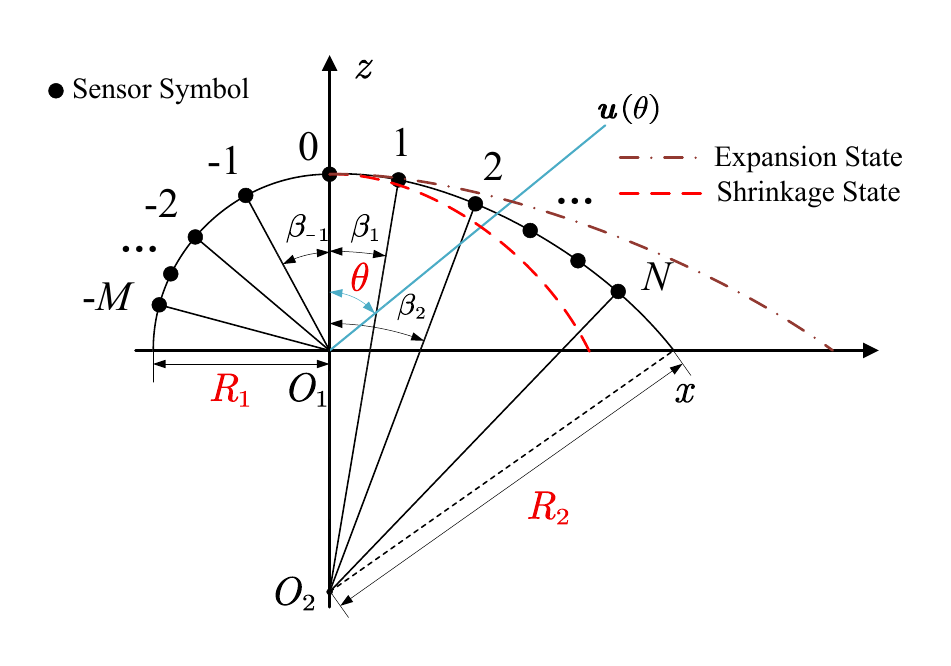}}
	\caption{ The configuration of flexible conformal array with two controlling parameters }
	\label{ring_array}
\end{figure}

SMWA will change wing shape according to different combat scenarios, flight phases, and enemy aircraft types and specifications.
For example, when a radar needs to detect or evade a target, a larger or smaller array aperture is required accordingly, so the SMWA need control the wing for timely expansion or contraction, as shown in Fig.\ref{illustrate}. After being elongated and shortened, the spacing of the array element will become uneven, and specifically, the maximum pointing direction of the beam pattern is deflected and the sidelobe level rises sharply, which is shown at the bottom of Fig.\ref{illustrate}. The corresponding beampattern will inevitably deteriorate in transmission and reception patterns. However, the existing RAB method has serious performance losses when dealing with ACP errors of FCA. Therefore, we need to propose a RAB method for the presence of ACP errors so that FCA can work properly.
In view of this, we present a FCA wing geometry as shown in Fig. \ref{ring_array}. The elements are uniformly distributed over two splice rings with half-wavelength inter-element spacing (in this case it means geodesic distance), and the elements have the same isotropic pattern with known positions. The ACPs are the two radiuses of the splice rings (i.e., $ R_1 $ and $ R_2 $), illustrating that the radiuses can be increased or decreased in the operation of the deformable surface. This model can be used to simulate airfoil adjustments such as camber deformation and thickness corresponding to SMWA \cite{CHU2022220}.  

Further, suppose that the array of $ M+N $ sensors receives signals from multiple narrowband sources. Let $ S=M+N$ represents the total number of all elements. The array observation vector $ \mathbf{x}\left( k \right) \in \mathbb{C} ^{S\times1} $ at time $ k $ can be modeled as
\begin{equation}
\label{1}
\mathbf{x}\left( k \right) =\mathbf{x}_s\left( k \right) +\mathbf{x}_i\left( k \right) +\mathbf{x}_n\left( k \right), 
\end{equation}
where $ \mathbf{x}_s\left( k \right) $, $ \mathbf{x}_i\left( k \right) $, and $ \mathbf{x}_n\left( k \right)  $ are the statistically independent components of the desired signal, interference, and noise, respectively. Another, consider the wavefield generated by one desired and $ L-1 $ interference sources located at  $ \left\{\theta _1,\theta _2,\cdots ,\theta _L\right\}  $, so that the desired signal $ \mathbf{x}_s\left( k \right)  $ and interference $ \mathbf{x}_i\left( k \right) $ can be written as 
\begin{align}
\mathbf{x}_s\left( k \right) =&\boldsymbol{C}\mathbf{a}\left( \theta _1 \right) s_1\left( k \right) 
\\
\mathbf{x}_i\left( k \right) =&\sum\nolimits_{j=2}^\mathrm{L}{\boldsymbol{C}\mathbf{a}\left( \theta _j \right) s_j\left( k \right)},
\label{2}	 
\end{align}
where $ s_j\left( k \right) $ is the sample transmitted by $ j^{th} $ source and $ \mathbf{a}\left( \theta_j \right) \in \mathbb{C} ^{S\times 1} $ is the complex SV associated with the source signal located at $ {\theta_j } $. $ \boldsymbol{C} \in \mathbb{C} ^{S\times S}  $ is mutual coupling matrix (MCM), and $ \mathbf{x}_n\left( k \right) \in  \mathbb{C} ^{S\times 1} $ is the complex vector of ambient channel noise samples that subject to Gaussian distribution in general. Then
\begin{equation}\label{3}
\mathbf{x}\left( k \right) =\sum\nolimits_{j=1}^\mathrm{L}{\boldsymbol{C}\mathbf{a}\left( \theta _j \right) s_j\left( k \right) +\mathbf{x}_n\left( k \right)}.
\end{equation}
Equation (\ref{3}) can be simplified by defining the effective array SV as \cite{4020392}
\begin{equation}\label{3+1}
\mathbf{\tilde{a}}\left( \theta _j \right) =\boldsymbol{C}\mathbf{a}\left( \theta _j \right). 
\end{equation}
Note that the matrix $ \boldsymbol{C} $ can be estimated by array calibration methods in \cite{7877862} or directly measuring the effective array pattern \cite{60990}, as it assumed to be independent of the DOA of signal \cite{509886}. Due to the flexibility of the FCA, the MCM varies with each deformation operation and should be considered in the algorithm. However, it is not convenient to measure MCM (e.g. full-wave simulation software HFSS) timely. Thus, a trade-off method is adopted to maintain the adaptivity of our approach. 
The mutual coupling between elements weakens with increasing spacing, and when the spacing is sufficiently large, the effect can be ignored \cite{509886}. Therefore, we only need to calculate it in small spacing state (e.g., when element spacing less than $ 2\lambda $, where $ \lambda $ is the wavelength of radiation frequency), and assume $ \boldsymbol{C}=\boldsymbol{I} $ ($ \boldsymbol{I} $ is the identity matrix) in other cases. 
Accordingly, each MCM of several selected states can be measured offline, and then approximated MCM will be called in real time for ABF processing. In this paper, the MCM is estimated using HFSS with a carefully selected set of ACPs. 

The typical adaptive beamformer output is given by \cite{0471733482}
\begin{equation}
y\left( k \right) =\mathbf{w}^H\mathbf{x}\left( k \right) 
\label{4}
\end{equation}
where $ \mathbf{w}=\left[ w_1,\cdots ,w_{S} \right] ^T\in \mathbb{C} ^{S\times1} $ is the beamformer weighted vector. 

Considering the mutual coupling effect, $ \mathbf{w} $ can be obtained by maximizing the output SINR, which is known as the Capon beamformer\cite{0471733482},
\begin{equation}
\mathbf{w}_{\mathrm{opt}}=\frac{\mathbf{R}_{i+n}^{-1}\mathbf{\tilde{a}}\left( \theta _1 \right)}{\mathbf{\tilde{a}}^H\left( \theta _1 \right) \mathbf{R}_{i+n}^{-1}\mathbf{\tilde{a}}\left( \theta _1 \right)}
\label{7}
\end{equation}
where $ \mathbf{\tilde{a}}\left( \theta _1 \right) $ is the presumed efficient array SV of $ \theta_1 $ direction, $ \mathbf{R}_{i+n}\triangleq E\{(\mathbf{x}_i\left( k \right) +\mathbf{x}_n\left( k \right) )(\mathbf{x}_i\left( k \right) +\mathbf{x}_n\left( k \right) )^H\}\in \mathbb{C} ^{ S  \times  S } $ is the interference-plus-noise covariance matrix (INCM). In practice, $ \mathbf{R}_{i+n} $ is usually unavailable even in signal-free applications, so it is replaced by the sample covariance matrix (SCM), i.e., $ \mathbf{\hat{R}}={{1}/{K}}\sum\nolimits_{k=1}^K{\mathbf{x}\left( k \right) \mathbf{x}^H\left( k \right)} $ with $ K $ snapshots, and the corresponding adaptive beamformer is the so-called sample matrix inversion (SMI) beamformer. 

Compared to the regular linear or planner array, the beamforming  performance of FCA is inherently affected by a considerable facets such as ACP errors and SV mismatches. This impart complexity and susceptibility into RAB by the extra flexibility. Consider the practical application scenario, there always exists the mismatch of SV caused by the error of CP during each deformable operations of the FCA.
According to the specific FCA model shown in Fig.\ref{ring_array}, the SV can be denoted as,
\begin{equation}
\mathbf{a}\left( \theta _i,r_1,r_2 \right) =\left[ \begin{matrix}
e^{jk\boldsymbol{p}_1\left( r_1,r_2  \right) \cdot \boldsymbol{u}\left( \theta _i \right)}&		\cdots&		e^{jk\boldsymbol{p}_S\left( r_1,r_2  \right) \cdot \boldsymbol{u}\left( \theta _i \right)}\\
\end{matrix} \right] ^T
\label{8}
\end{equation}
Here, $ r_1 $ and $ r_2 $ are two independent random variables representing the true values of $ R_1 $ and $ R_2 $, which fluctuate over small ranges. $ {\boldsymbol{u}}( \theta _i) $ denotes the unit vector of the impinging direction of the source in $ \theta _i $, i.e., $ \boldsymbol{u}\left( \theta _i \right) =\left[ \begin{matrix}
	\sin \theta _i&		\cos \theta _i\\
\end{matrix} \right] ^T $.
$  \boldsymbol{p}_1, \cdots , \boldsymbol{p}_S$ denote the nominal location of each element respectively, i.e. $ \boldsymbol{p}_s=r_1\left[ \begin{matrix}
	\sin\mathrm{(}\beta _s)&		\cos\mathrm{(}\beta _s)\\
\end{matrix} \right] ^T$, if $s=-1,\cdots ,-M $, and $ \boldsymbol{p}_s=r_1\left[ \begin{matrix}
0&		1\\
\end{matrix} \right] ^T+r_2\left[ \begin{matrix}
\sin\mathrm{(}\beta _s)&		\cos\mathrm{(}\beta _s)-1\\
\end{matrix} \right] ^T$, if $s=1,\cdots ,N $, and $ \beta _s$ represents the angle at the center of the circle $ O_1 $ or $ O_2 $, which can determine the position of each element. Usually, $ \beta _s=s\lambda /\left( 2R_1 \right) ,s=-1,\cdots ,-M$, and $\beta _s=s\lambda /\left( 2R_2 \right) ,s=1,\cdots ,N $, where $ R_1 $ and $ R_2 $ are the initial values. 

Another, we assume $ \left| r_1-\bar{r}_1 \right|\leqslant l_1,\left| r_2-\bar{r}_2 \right|\leqslant l_2 $. $ \bar{r}_1 $ and $ \bar{r}_2$ are the presumed values of ACPs, $ l_1 $ and $ l_2 $ are two real numbers representing the maximum errors of ACPs. In practice, the true values of ACPs cannot be measured precisely. In addition, the information at the arrival angle of the signal of interest (SOI) may be imprecise, and the knowledge of the array location maybe inaccurate or unreliable due to array calibration errors, vibrations and sensory errors. In particular, the SMI mismatch caused by a small snapshot should also be considered. Thus, from a holistic perspective, a more comprehensive optimal RAB problem for FCA is described as
\begin{equation}\label{19}
\left\{ \begin{array}{l}
\underset{\mathbf{w}}{\min}\,\,\mathbf{w}^H\mathbf{R}_{i+n}\mathbf{w}\,\,\\
s.t.\,\,\,\left| \mathbf{w}^H\mathbf{a}\left( \theta ,r_1,r_2 \right) \right|\geqslant 1,\left| \theta -\theta _1 \right|\leqslant \delta _{\theta}\\
\qquad    \left| r_1-\bar{r}_1 \right|\leqslant l_1,\left| r_2-\bar{r}_2 \right|\leqslant l_2\\
\qquad \mathbf{a}\left( \theta ,r_1,r_2 \right) \in \mathcal{A} \left( \mathbf{a}\left( \theta,r_1,r_2 \right) \right) \,\,\\
\end{array} \right. 
\end{equation} 
Here, $ \delta _{\theta} $ is a small number that represents the error in the DOA.
Due to the perturbation of the array elements induced by the above reasons, the actual signal SV usually belongs to the uncertainty set $ \mathcal{A} \left( \mathbf{a}\left( \theta ,r_1,r_2 \right) \right) \subset \mathbb{C} ^{S\times 1 }$, which is defined as
\begin{equation}\label{20}
\mathcal{A} \left( \mathbf{a}\left( \theta ,r_1,r_2\right) \right) =\left\{ \boldsymbol{c}\left| \boldsymbol{c}=\mathbf{a}\left( \theta ,r_1,r_2 \right) +\boldsymbol{e},\left\| \boldsymbol{e} \right\|_2 \leqslant \varepsilon \right. \right\}
\end{equation}
where $ \varepsilon >0 $ is a known constant which is bounded by the 2-norm of SV distortion. If we know some prior information about SV mismatch, (\ref{20}) can be reformulated as an ellipsoid uncertainty set, for more details see \cite{Lorenz1420809}.



\section{THE PROPOSED METHOD}
\label{pro}

For regular structure arrays, there are many approaches to dealing with the RAB problems which are discussed in \cite{4682560},\cite{1166614}, and\cite{8448458}. However, the performance of the aforementioned algorithms will degrade when they are applied to FCA due to its unique structure and radiation properties. Therefore,  a RAB algorithm based on extended multi-domain INCM reconstruction and SV estimation is proposed for FCA to compensate the errors of ACPs in this paper.
\subsection{ INCM reconstruction}
\begin{figure}[htp]
	\centering
	\subfigure {\includegraphics[width=0.6\textwidth]{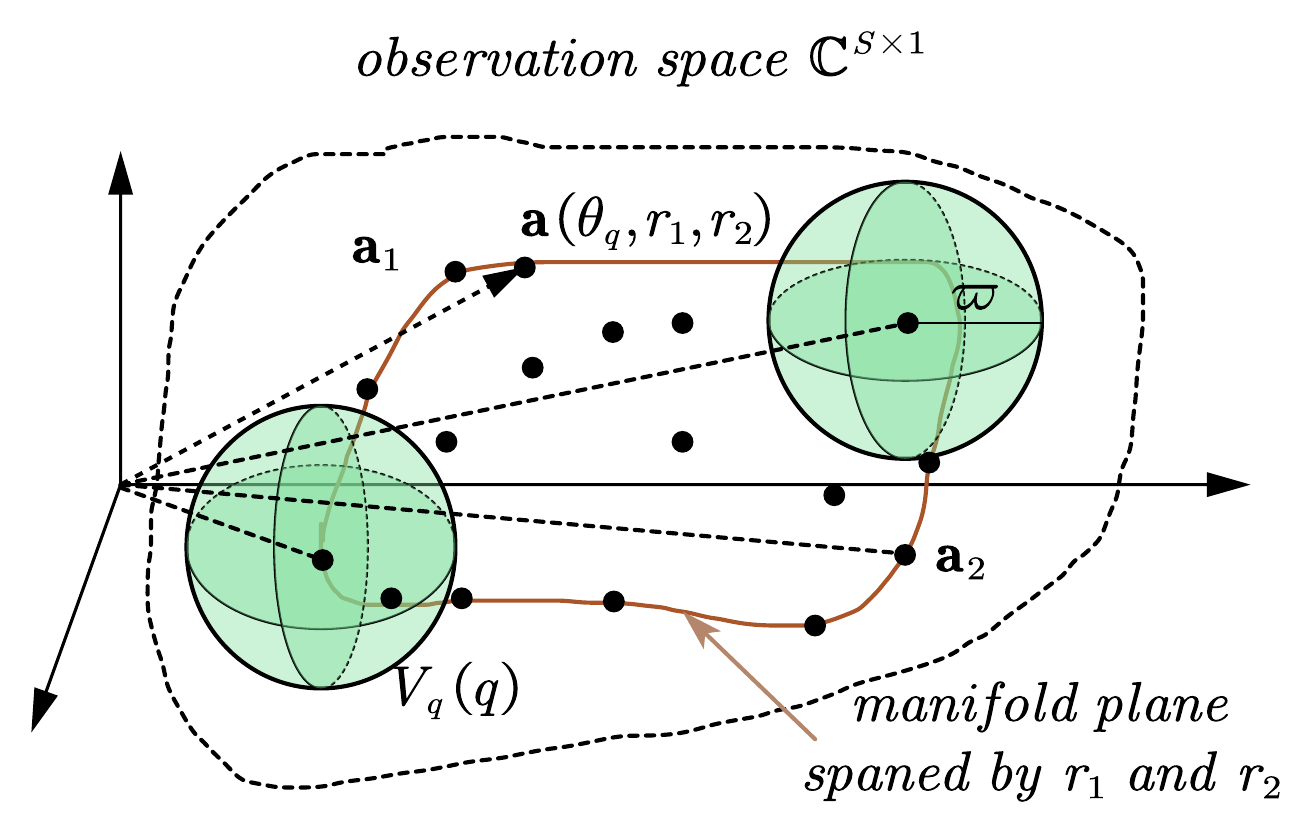}}
	\caption{ Graphical representation of the array manifold cube in $ \mathbb{C} ^{S\times 1 }$, $ \mathbf{a}_1 $ and $ \mathbf{a}_2 $ are the representations of possible SV which are denoted by carbon spots.}
	\label{manifold}
\end{figure}

In general, the INCM is reconstructed by calculating the integral of the spatial spectrum distribution over all possible directions that refer to the spatial domain except the hypothetical direction (SOI spatial direction) \cite{6180020}. For FCA, this domain is extended to an extra parameter domain that is different from the algorithms proposed in \cite{6180020} and \cite{7018037}. 
In detail, considering the mutual coupling effect, the INCM represented by $ \hat{\boldsymbol{R}}_{i+n} $ is reconstructed as
\begin{equation}\label{23}
\hat{\boldsymbol{R}}_{i+n}=\hat{\sigma}_{n}^{2}\mathbf{I}+\sum\nolimits_{q=1}^{T-1}{\iiint\limits_{V_{\mathbf{a}}\left( q \right)}{\frac{\boldsymbol{C}\mathbf{aa}^H\boldsymbol{C}^H}{\mathbf{a}^H\boldsymbol{C}^H\hat{\boldsymbol{R}}^{-1}\boldsymbol{C}\mathbf{a}}}d\mathbf{a}}
\end{equation} 
where $ \hat{\sigma}_{n}^{2} $ is the estimated noise power which usually equals to the minimal eigenvalue of $\hat{\boldsymbol{R}}  $. $ V_{\mathbf{a}}\left( q \right) $ is the uncertainty domain in $ \mathbb{C} ^{S\times 1 } $ corresponding to the $ q $-th interference, denoted as 
\begin{equation}\label{24}
\begin{aligned}
V_{\mathbf{a}}\left( q \right) &=\left\{ \boldsymbol{a}\mid \left\| \boldsymbol{a}-\mathbf{a}\left( \theta _q,r_1,r_2 \right) \right\|_2 \leqslant \varpi , \right.\\
&\qquad \left. \left| r_1-\bar{r}_1 \right|\leqslant l_1,\left| r_2-\bar{r}_2 \right|\leqslant l_2,\theta _q\in \Theta _q \right\}\\
\end{aligned}
\end{equation}
where $ \theta _q $ is the presumed direction of $ q$-th interference, and $ \Theta _q $ is the corresponding range induced by the inaccurate estimation of the direction. $ \varpi $ constrains the possible SV with a sphere in high dimensional complex space $ \mathbb{C} ^{S\times 1 } $. Fig.\ref{manifold} shows an intuitive explanation of the $ V_{\mathbf{a}}\left( q \right) $ and the corresponding point in $ \mathbb{C} ^{S\times 1 }  $ space.
Note that, although it is easy to find the parallel and orthogonal component of a specified SV (i.e., divide $ V_{\mathbf{a}}\left( q \right) $ into many vector sets) in complex space, it is difficult to calculate the integral in that space. Thus we choose to approximate (\ref{23}) by calculating the discrete sum as
\begin{equation}\label{25}
\hat{\boldsymbol{R}}_{i+n}=\hat{\sigma}_{n}^{2}\mathbf{I}+\frac{4l_1l_2}{S_{r_1}S_{r_2}}\sum_{q=1}^{T-1}{\frac{L_{\varTheta}}{S_{\Theta _q}}\sum_{r_1}{\sum_{r_2}{\sum_{\theta}{\frac{\boldsymbol{C}\mathbf{aa}^H\boldsymbol{C}^H}{\mathbf{a}^H\boldsymbol{C}^H\hat{\boldsymbol{R}}^{-1}\boldsymbol{C}\mathbf{a}}}}}}
\end{equation}
where $ {S_{r_1},S_{r_2},S_{\Theta _q}} $ represents the corresponding sampling number in parameter range, respectively. $L_{\varTheta}$ represents the length of interval $  \Theta _q$. 

\subsection{SV estimation using Alternative optimization}
Assume the real SV can be denoted as
\begin{equation}\label{appp1}
\mathbf{a}=\bar{\mathbf{a}}\left( \theta _1,r_1,r_2 \right) +\mathbf{e}_{\bot}
\end{equation}
 where $ \mathbf{\bar{a}}\left( \theta _1 ,r_1,r_2\right)  $ is the presumed SV of SOI with parameter $ \left( \theta _1 ,r_1,r_2\right)   $, which is a {function of ACPs} due to the special feature of FCA. $ \mathbf{e}_{\bot} $ is one component of the mismatch vector $ \mathbf{e} $, which is orthogonal to $ \mathbf{\bar{a}} $. 
Then, we estimate the SV of SOI by solving the following optimization problem,
\begin{equation}\label{26}
\left\{ \begin{aligned}
\underset{r_1,r_2,\mathbf{e}_{\bot}}{\min}&\,\,\left( \bar{\mathbf{a}}\left( \theta _1 ,r_1,r_2\right) +\mathbf{e}_{\bot} \right) ^H\grave{\boldsymbol{R}}\left( \bar{\mathbf{a}}\left( \theta _1 ,r_1,r_2\right) +\mathbf{e}_{\bot} \right)\\
s.t.\quad &\bar{\mathbf{a}}^H\mathbf{e}_{\bot}=0,\bar{\mathbf{a}}\in V_{\mathbf{a}}\left( 1 \right)\\
&\left( \bar{\mathbf{a}}+\mathbf{e}_{\bot} \right) ^H\grave{\boldsymbol{R}}\left( \bar{\mathbf{a}}+\mathbf{e}_{\bot} \right) \leqslant \bar{\mathbf{a}}^H\grave{\boldsymbol{R}}\bar{\mathbf{a}}\\
\end{aligned} \right. 
\end{equation}

For simplicity, $ \grave{\boldsymbol{R}}=\boldsymbol{C}^H\hat{\boldsymbol{R}}^{-1}\boldsymbol{C} $, and the dependence of various vectors and matrices on the variable will be dropped and only explicitly mentioned when needed (in this case $\bar{\mathbf{a}} $ denotes $ \bar{\mathbf{a}}\left( \theta _1,r_1,r_2 \right) $). The other component is parallel to $ \mathbf{\bar{a}} $ which is omitted here because beamforming quality will not be affected by a scaled copy of it. 
Then, the equality constraint is introduced to maintain orthogonality between them, and the inequality constraint is aimed at preventing the actual SV from converging to any interference in the range of SOI.

Note that (\ref{26}) is a multivariate nonconvex optimization problem, using the idea of alternating direction multipliers (ADMM) \cite{Boyd8186925}, (\ref{26}) can be factorized into two alternative optimization problems (see Appendix \ref{app1} for detailed deduction):
\begin{equation}\label{27}
p_{1}^{\left( m \right)}:\left\{ \begin{aligned}
&\underset{\mathbf{e}_{\bot}}{\min}\,\,( \bar{\mathbf{a}}^{\left( m-1 \right)}+\mathbf{e}_{\bot} ) ^H\grave{\boldsymbol{R}}( \bar{\mathbf{a}}^{\left( m-1 \right)}+\mathbf{e}_{\bot} )\\
&s.t.\quad {\mathbf{e}^H}_{\bot}\bar{\mathbf{a}}^{\left( m-1 \right)}=0,\\
&\qquad( \bar{\mathbf{a}}^{\left( m-1 \right)}+\mathbf{e}_{\bot} ) ^H\grave{\boldsymbol{R}}( \bar{\mathbf{a}}^{\left( m-1 \right)}+\mathbf{e}_{\bot} )\\
&\qquad \qquad\qquad\quad\leqslant ( \bar{\mathbf{a}}^{\left( m-1 \right)} ) ^H\grave{\boldsymbol{R}}\bar{\mathbf{a}}^{\left( m-1 \right)}\\
\end{aligned} \right. 
\end{equation}
and
\begin{equation}\label{28}
p_{2}^{\left( m \right)}:\left\{ \begin{aligned}
&\underset{r_1,r_2}{\min}\,\,( \bar{\mathbf{a}}+\mathbf{e}_{\bot}^{\left( m \right)} ) ^H\grave{\boldsymbol{R}}( \bar{\mathbf{a}}+\mathbf{e}_{\bot}^{\left( m \right)} )\\
&s.t.\quad \bar{\mathbf{a}}^H\mathbf{e}_{\bot}^{\left( m \right)}=0,\left| r_1-\bar{r}_1 \right|\leqslant l_1,\left| r_2-\bar{r}_2 \right|\leqslant l_2\\
&\qquad ( \bar{\mathbf{a}}+\mathbf{e}_{\bot}^{\left( m \right)} ) ^H\grave{\boldsymbol{R}}( \bar{\mathbf{a}}+\mathbf{e}_{\bot}^{\left( m \right)} ) \leqslant \bar{\mathbf{a}}^H\grave{\boldsymbol{R}}\bar{\mathbf{a}}\\
\end{aligned} \right. 
\end{equation}
where $ m=1,2,3\cdots $ is the iterative index. Note that, $ p_{1}^{\left( m \right)} $ is a QCQP problem if we fix $ \bar{\mathbf{a}}$, which can be solved using convex optimization software, such as CVX \cite{cvx}. Then we focus on $ p_{2}^{\left( m \right)} $, because the optimization variables (i.e., $ r_1$ and $r_2 $) are in the exponent part. It is difficult to deal with this nonconvex optimization problem. So we resort to \textsl{inner point method} \cite{KuhnHaroldW}.
By this we can alternatively solve the two optimization problems equivalently in order to solve (\ref{26}) .

For the convergence of the proposed alternative optimization method, we know that $\bar{\mathbf{a}}+\mathbf{e}_{\bot} $ is convergent because $ \mathbf{e}_{\bot} $ is constrained by quadratic constraints and the norm-2 of $ \bar{\mathbf{a}} $ is constant.

After $ M_r $ iterations, the optimal result of  $ p_{1}^{\left( m \right)} $ and $ p_{2}^{\left( m \right)} $ converge to $\mathbf{e}_{\bot}^{ *} $ and $ \bar{\mathbf{a}}^{ *} $, respectively. Then substitute them and (\ref{25}) into the Capon beamformer (\ref{7}), we obtain the optimal  beamformer as
\begin{equation}\label{47}
\mathbf{w}^{md}=\frac{\hat{\boldsymbol{R}}_{i+n}^{-1}\boldsymbol{C}\left( \bar{\mathbf{a}}^*+\mathbf{e}_{\bot}^{*} \right)}{\left( \bar{\mathbf{a}}^*+\mathbf{e}_{\bot}^{*} \right) ^H\boldsymbol{C}^H\hat{\boldsymbol{R}}_{i+n}^{-1}\boldsymbol{C}\left( \bar{\mathbf{a}}^*+\mathbf{e}_{\bot}^{*} \right)}
\end{equation}
The proposed algorithm is summarized in Algorithm \ref{algorithm1}.
\begin{algorithm}[htb]
	\caption{Proposed robust beamforming algorithm for FCA.}
	\label{algorithm1}
	\begin{algorithmic}[1]
		\Require\\
		
		$ \hat{\boldsymbol{R}} $ ;$\mathbf{\bar{a}}^{\left( 0 \right)} $; $\varpi $;
 $ \bar{r}_1,\bar{r}_2,l_1,l_2 $;
$\boldsymbol{C}$;\\
		The parameter sequence of barrier weight: $\varepsilon ^{\left( k \right)}$;\\		
		The iterative stopping condition: $ \tau _x,\tau _y,\tau _{\boldsymbol{e}},D  $.
		\Ensure
$ \bar{\mathbf{a}}^{ *} $ ; $\mathbf{e}_{\bot}^{ *}  $; $ \hat{\boldsymbol{R}}_{i+n} $.
		\State Determine the discrete sampling set of parameters corresponding to each sources (totally $ L $ terms), i.e., $ \left( \theta ,r_1,r_2 \right) \in \mathop {\underbrace{\left\{ \theta _1,\cdots \right\} }} \limits_{S_{\theta}}\times \mathop {\underbrace{\left\{ \bar{r}_1,\cdots \right\} }} \limits_{S_{r_1}}\times \mathop {\underbrace{\left\{ \bar{r}_2,\cdots \right\} }} \limits_{S_{r_2}} $, and to reconstruct: $ \boldsymbol{\hat{R}}_{i+n} $ using (\ref{25});
		\label{a1}
		\State initialize $ m=1 $ and $\mathbf{\bar{a}}= \mathbf{\bar{a}}^{\left( 0 \right)}  $;
		\State Solving the optimization problem $ p_{1}^{\left( m \right)} $ using CVX, and obtaining the optimal result of $ \mathbf{e}_{\bot}^{\left( m \right)} $;
		\label{a2}
		\State Set $ \mathbf{e}_{\bot}=\mathbf{e}_{\bot}^{\left( m \right)}   $; and solve the optimization problem $ p_{2}^{\left( m \right)} $ by \textsl{inner point method} and \textsl{Newton's method} (using (\ref{ap11})) to obtain $ x^{(m)},y^{(m)} $ and $ \bar{\mathbf{a}}^{\left( m \right)}  $;
		\label{a3}
		\State \textbf{If }$ \left| x^{(m)}-x^{(m-1)} \right|\leqslant \tau _x,\left| y^{(m)}-y^{(m-1)} \right|\leqslant \tau _y,\left\| \mathbf{e}_{\bot}^{\left( m \right)}-\mathbf{e}_{\bot}^{\left( m-1 \right)} \right\|_2 \leqslant \tau _{\boldsymbol{e}} $; or $ m>D $, 
		
		go to step \ref{a7}.\\
		
		\textbf{else} 
		
		set $ m:=m+1 $ and go to step \ref{a2};
		\label{a4}
		\\		
		\Return $ \boldsymbol{\hat{R}}_{i+n} $, $\bar{\mathbf{a}}^{ *}  $, $\mathbf{e}_{\bot}^{ *}$.
		\label{a7}
	\end{algorithmic} 
\end{algorithm}

%


\subsection{COMPLEXITY ANALYSIS}
\label{comp}
Suppose we only consider the numbers of multiplication operations (MOs). Table \ref{addt.1} illustrates the required number of MOs for several mathematical operations (see \cite{8931778}). 
\begin{table}[!tbp]	
	\caption{Required number of MOs for typical mathematical operations}
	\centering
	\begin{tabular}{|l|c|}
		\hline
		Operation              & Required number of MOs \\ \hline
		Exponential($ e^x $)         & 15                     \\ \hline
		Sin($ x $)                 & 7                      \\ \hline
		Cos($ x $)                 & 8                      \\ \hline
		Complex multiplication & 6             \\ \hline
		Matrix multiplication($ \boldsymbol{A}_{m\times n}\boldsymbol{B}_{n\times d} $) & $ m\cdot n\cdot d $             \\ \hline
		Matrix inversion($ \boldsymbol{A}_{n\times n} $) & $ \mathcal{O} \left( n^3 \right)  $             \\ \hline
		SVD decomposition &$ \mathcal{O} \left( n^3 \right)  $             \\ \hline
		QCQP problem($ n $ is order)\cite{6891348} &$ \mathcal{O} \left( n^3 \right)  $             \\ \hline           
	\end{tabular}
	\label{addt.1}
\end{table}
\begin{enumerate}[]
	\item INCM reconstruction: Assuming the iteration number of SV estimation is $ Q_s $, for one iteration of it, we have:
	\begin{itemize}
		\item $ \boldsymbol{\hat{R}}_{i+n} $ needs $4LS^2S_{\theta}S_{r_1}S_{r_2}$ MOs.
		\item To estimate $ \bar{\mathbf{a}}^{\left( m \right)}  $, we need $ \mathcal{O} \left( S^3 \right)  $  MOs.
		\item To estimate $ \mathbf{e}_{\bot}^{\left( m \right)} $ by solving QCQP, we need $ \mathcal{O} \left( S^3 \right)  $  MOs.
	\end{itemize}
	\item Hence, the total complexity for TS-SQP is
	\begin{equation}\label{65}
	\begin{cases}
	\mathcal{O} \left( S^3Q_s \right), \,\, if\,\, S\gg S_{ma}\\
	\mathcal{O} \left( S_{ma}^{3}S \right), \,\,if\,\,S_{ma}\gg S\\
	\end{cases}
	\end{equation}
	where we take the notation of $S_{ma} =max\{Q_s,S_{r_1},S_{r_2},S_{r_\theta}\}$.
\end{enumerate}
\section{SIMULATION RESULT}\label{simu}
Consider 23-element FCA ($ M=9,N=13 $, and $ R_1=500\left( mm \right) $, $ R_2=1000\left( mm \right) $) with fixed central angle $ \beta _{-1}=-\lambda/(2R_1)$ or $\beta _1 =\lambda/(2R_2) $, respectively, as shown in Fig.\ref{ring_array}. The desired direction is $ \theta _1=10\degree $. There are three equipowered interferers located at $ -20^{\degree} $, $ -10^{\degree} $ and $ 20^{\degree} $, and the signal noise ratio (SNR) is 30dB. The additive noise is modeled as a Gaussian zero-mean spatial and temporal white process.

To obtain the mutual coupling matrix $ \boldsymbol{C} $, the impedance matrix $ \boldsymbol{Z} $ is obtained using HFSS software and then calculated by
\begin{equation}\label{s1}
\boldsymbol{C}=\left( Z_A+Z_T \right) \left( \boldsymbol{Z}+Z_T\boldsymbol{I} \right) ^{-1}.
\end{equation}
where $ Z_A+Z_T=2Re\left\lbrace diag\left(\boldsymbol{Z}\right) \right\rbrace $, $Re\left\lbrace \bullet \right\rbrace$ is the real part of brace. The angular region of one desired signal is set to be $ \varTheta =\left[ \theta _1-5{\degree},\theta _1+5{\degree} \right] $, and the same range is used by each interference, while the complementary angular sectors of $ \varTheta $ are $ \bar{\varTheta}=\left[ -90{\degree},\theta _1-5{\degree} \right] \cup \left[ \theta _1+5{\degree},90{\degree} \right] $ in order to implement classical INCM reconstruction beamformer. All integral operations in this paper are replaced by discrete sums and all angular sectors are uniformly sampled with the same angular interval $ 0.5\degree $. For each scenario, 100 Monte-Carlo runs are performed.

The performance of the proposed algorithm is compared with: diagonal loading method (DL-SMI) proposed in \cite{4020392}, eigenspace projection methods ( Eigenspace) in \cite{650188}, norm and uncertainty set constrained method (DCRCB) in \cite{0471733482} and the classical INCM reconstruction and SV estimation method (Reconstruct) in \cite{6180020}. 

For the proposed beamformer, we set $( \tau _x,\tau _y,\tau _{\boldsymbol{e}})=(0.01,0.1,0.1) $ and carefully design $ \epsilon ^k=\left( 100,\cdots ,1 \right) $ with 10 terms uniformly. Convex optimization toolbox CVX \cite{cvx} is used to solve optimization problems.
\subsection{Example1: Mismatch due to ACP Errors}
\begin{figure}[htbp]
	\centering
	\subfigure {\includegraphics[width=0.7\textwidth]{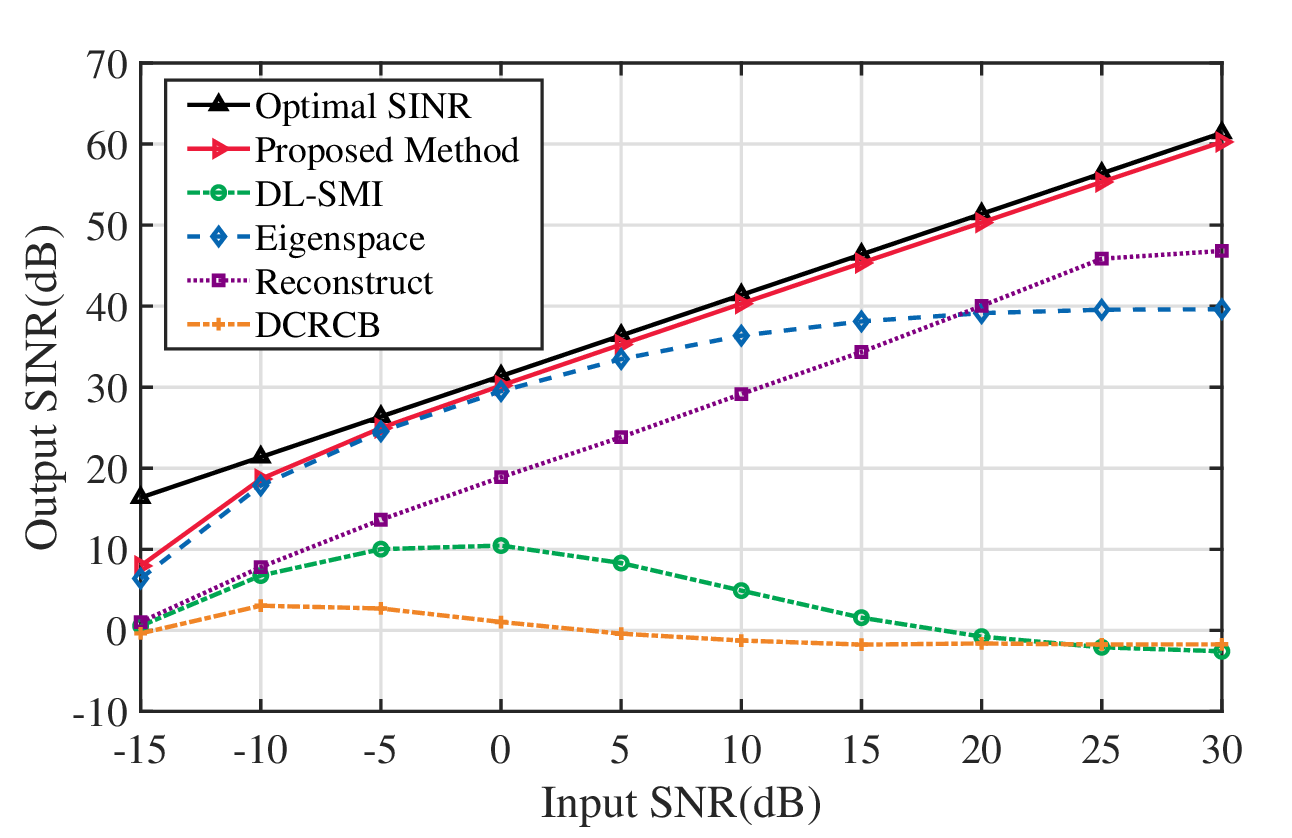}}
	\caption{ Output SINR versus the input SNR in case of ACP errors for the first scenario.}
	\label{f.2}
\end{figure}

In the first example, we simulate a scenario where the actual ACPs of the FCA are not exactly known and the mutual coupling effect is ignored. We set the fixed errors of ACPs, $ \varDelta R_1=-0.21\, mm  $, $ \varDelta R_2=3.7\,mm $. Fig.\ref{f.2} shows the output SINR curves of the tested methods versus the input SNR for the fixed snapshot $ K=2S=46 $. It shows that the proposed method achieves better performance than Eigenspace and INCM reconstruct methods, etc. It is worth noting that the curve of the Eigenspace beamformer is close to the proposed one at low SNR, but it suffers degradation in high input SNR conditions. Because the proposed method greatly benefits the highly accurate estimations of both the SV of SOI and INCM, its performance is very close to the optimal curve. Whereas the Reconstruct method suffers performance degradation because it does not consider this fact. Moreover, Fig.\ref{f.2} has shown that a small error of ACPs ($ \varDelta R_1=-0.21 \,mm $ compared to $ R_1=500\,mm $) induce a significant loss of output SINR, since there is a gap of 10dB approximately between the curves of the proposed and INCM reconstruction methods. Due to the existence of ACPs error, the DCRCB method can not fully cover the range of the real SV, so the performance is always poor. The performance of DL-SMI decreases with the increase of input SNR because it can not accurately estimate the loading factor and the SCM when SNR is high.

\begin{figure}[htbp]
	\centering
	\subfigure {\includegraphics[width=0.7\textwidth]{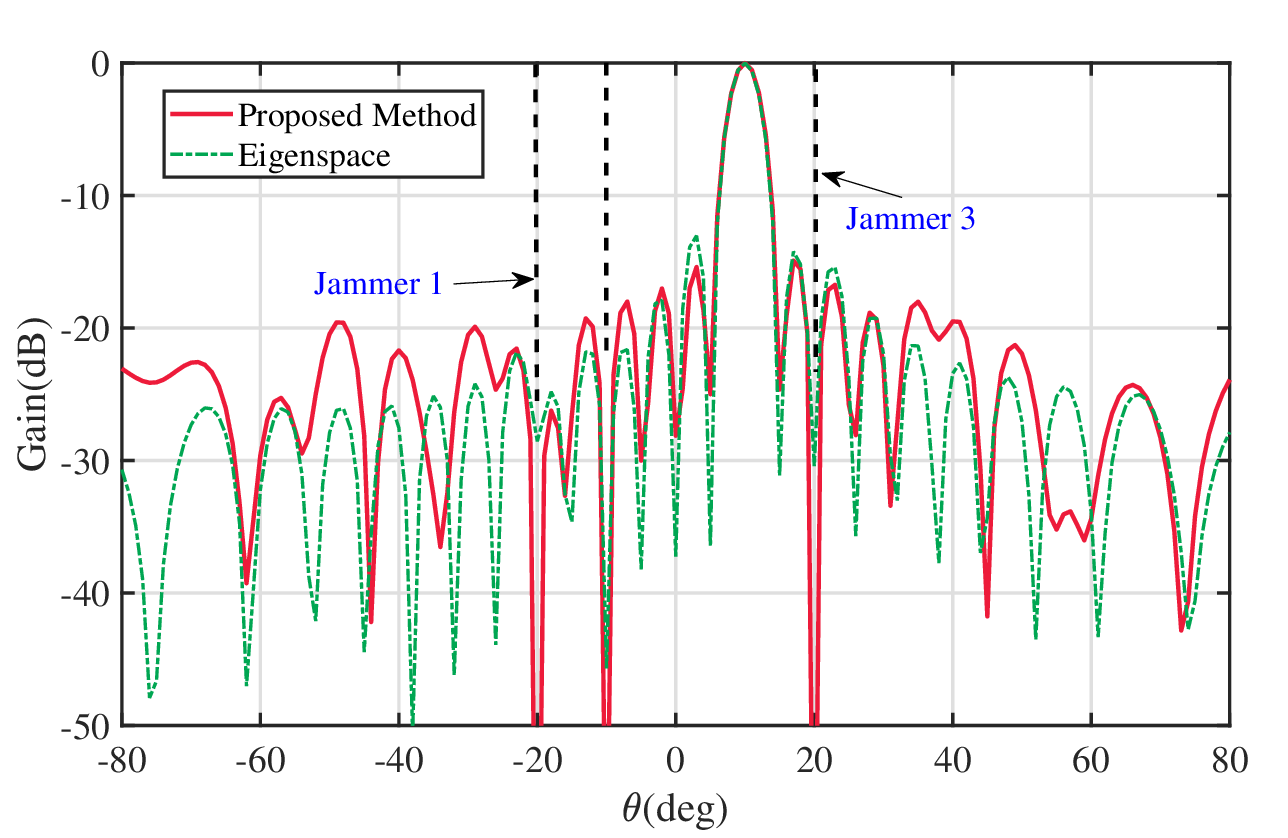}}
	\caption{ Steady state beampatterns in case of ACP errors for first scenario ( SNR=10dB ).}
	\label{f.4}
\end{figure}

In order to show the interference rejection performance of the proposed method intuitively, we compare our method with the Eigenspace method, which performs similarly to our proposed method. It is clear from Fig.\ref{f.4} that the proposed beamformer provides the highest desired signal beam at $ 10\degree $ and has the deepest nulls at DOAs (i.e., -10\degree, -20\degree, 20\degree) compared with the Eigenspace method. In addition, the proposed beamformer exhibits a better sidelobe suppression capability than the Eigenspace method.

\subsection{Example2: Mismatch due to Looking Direction Error and random ACPs errors with SV perturbation}
\begin{figure}[htbp]
	\centering
	\subfigure {\includegraphics[width=0.7\textwidth]{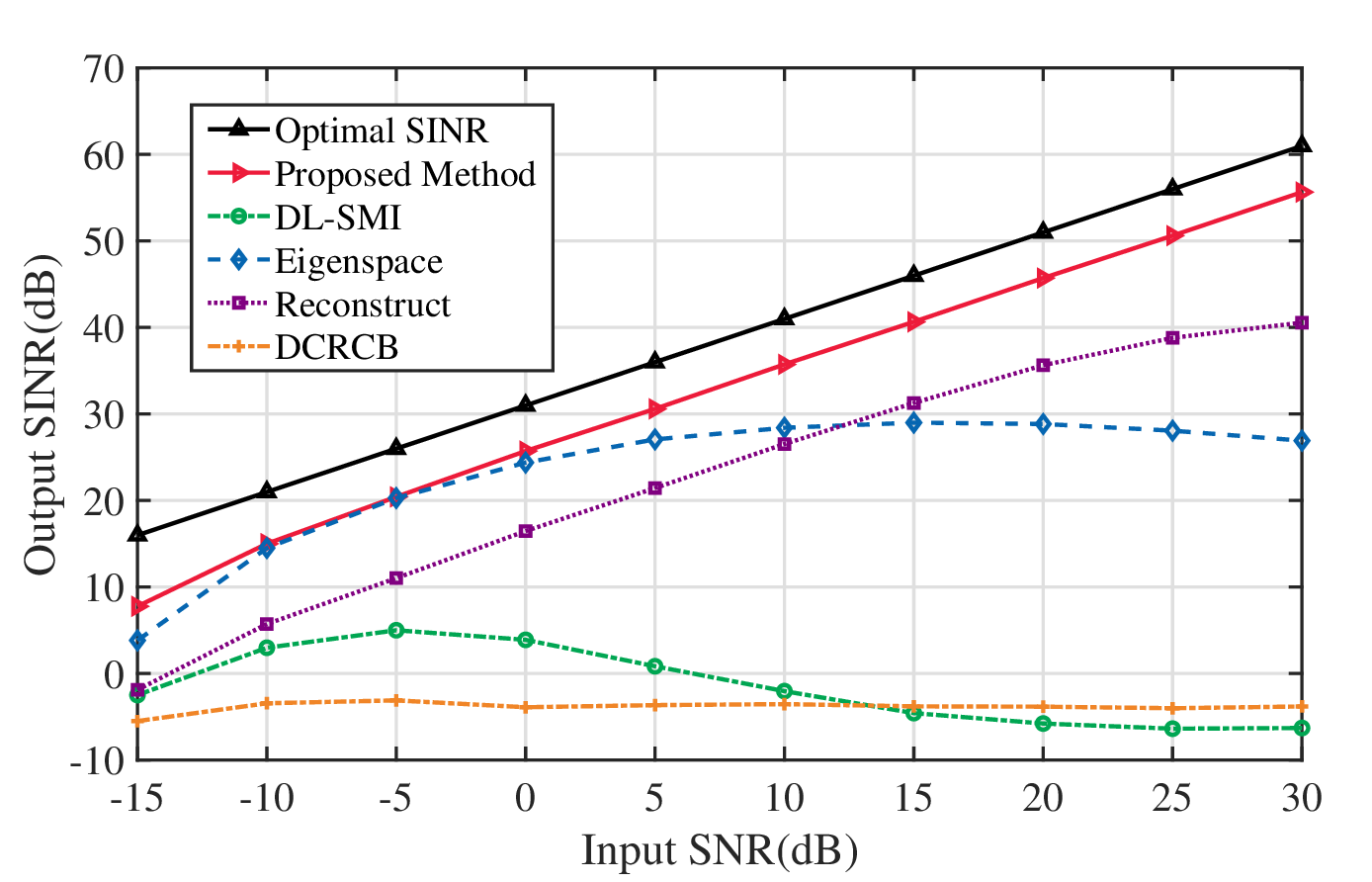}}
\caption{ Output SINR versus the input SNR for the second scenario.}
	\label{f.5}
\end{figure}

\begin{figure}[htbp]
	\centering
	\subfigure {\includegraphics[width=0.7\textwidth]{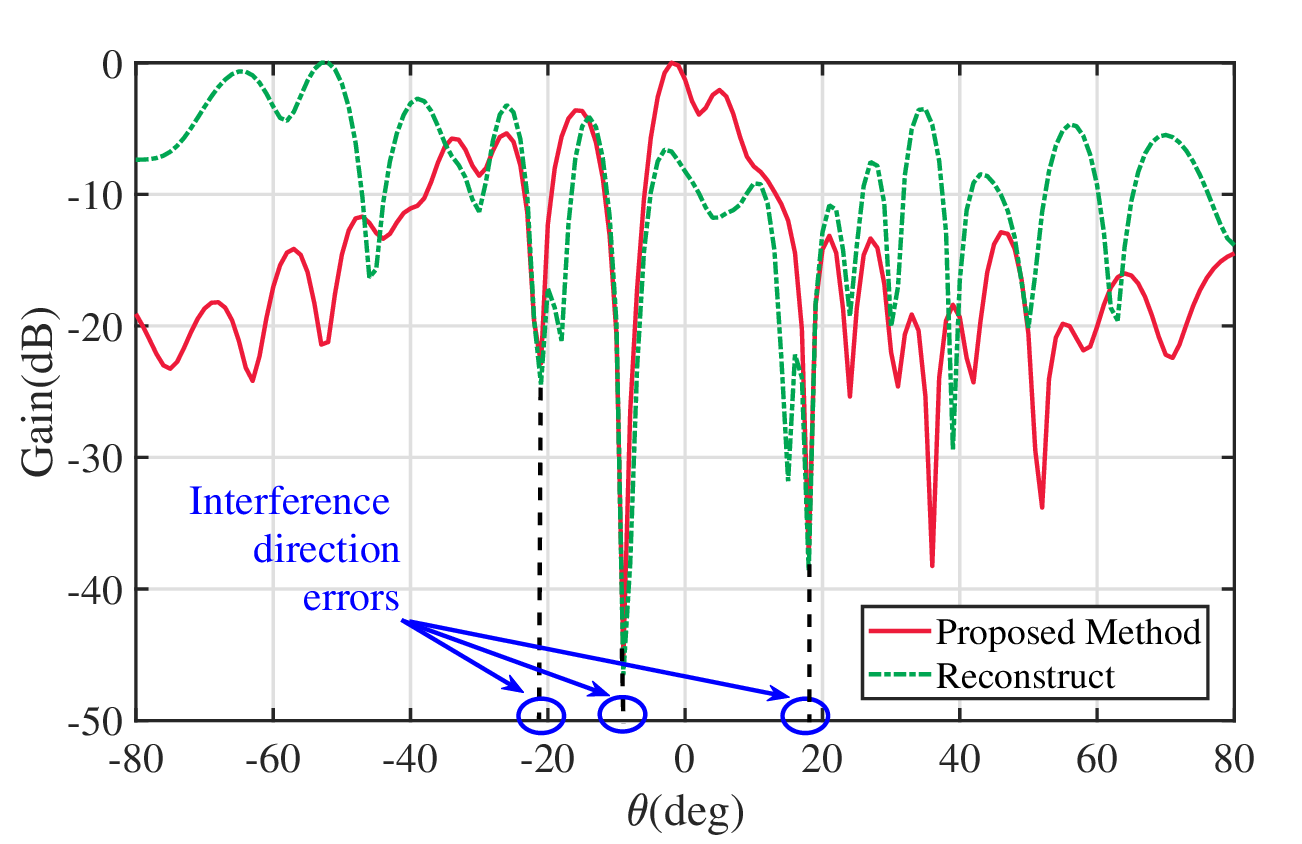}}
	\caption{ Steady state beampatterns for the second scenario ( SNR=10 dB ).}
	\label{f.7}
\end{figure}

In the second example, assume that the random direction errors of desired signal and interference are uniformly distributed in $ \left[ -5{\degree},5{\degree} \right] $ for each simulation, and associated with the random ACP errors of FCA which are uniformly distributed in $ \left[ -15,15 \right]\,mm $ of $ \Delta R_1 $ and $ \left[ -20,20 \right]\,mm $ of $ \Delta R_2 $. Fig.\ref{f.5} shows that the proposed method achieves better performance than others, but there is a 8 dB performance gap between the proposed method and the optimal one. The main reason is the mismatch of the estimated SV with the actual one in the resulting beamformer, which has a slight derivation in this comprehensive case. As a comparation, we compare our method with the Reconstruct method. It is also shown in quiescent beampatterns in Fig.\ref{f.7} that the maximum gain direction (i.e., 0\degree) deviates from the actual SOI (i.e., 10.5\degree in this case).


Moreover, Fig.\ref{f.7} shows the ability of the proposed method to have the deepest nulls at the actual interferer DOAs (i.e., 17.8\degree, -24.1\degree, -8.9\degree) which have small deviations from the presumed ones, associated with a wide gain range in the neighbor of the actual SOI direction. In addition, the proposed beamformer exhibits better sidelobe suppression capability than the conventional Reconstruct method in the case of hybrid looking direction error and ACPs errors.

\subsection{Example3:Impacts of mutual coupling effect associated with mixed mismatches}
\begin{figure}[htp]
	\centering
	\subfigure {\includegraphics[width=0.9\textwidth]{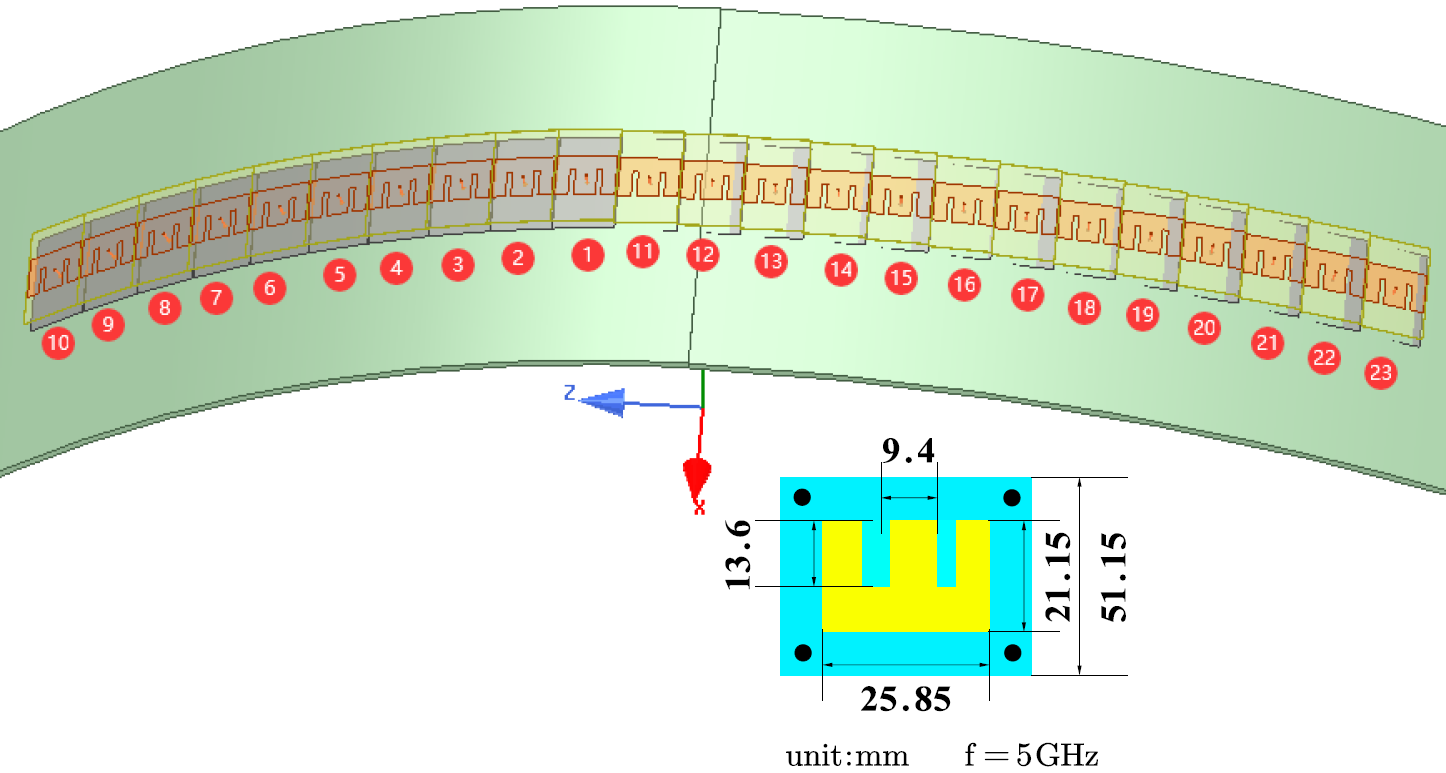}}
	\caption{ Geometry of a 23-element sectorial FCA (with view of the E-type antenna element on cylindrical surface, a substrate with $ \varepsilon _r=2.2 $ and a thickness of 1.57 mm is used, the work frequency is 5GHz }
	\label{f.11}
\end{figure}

\begin{figure}[htbp]
	\centering
	\subfigure {\includegraphics[width=0.65\textwidth]{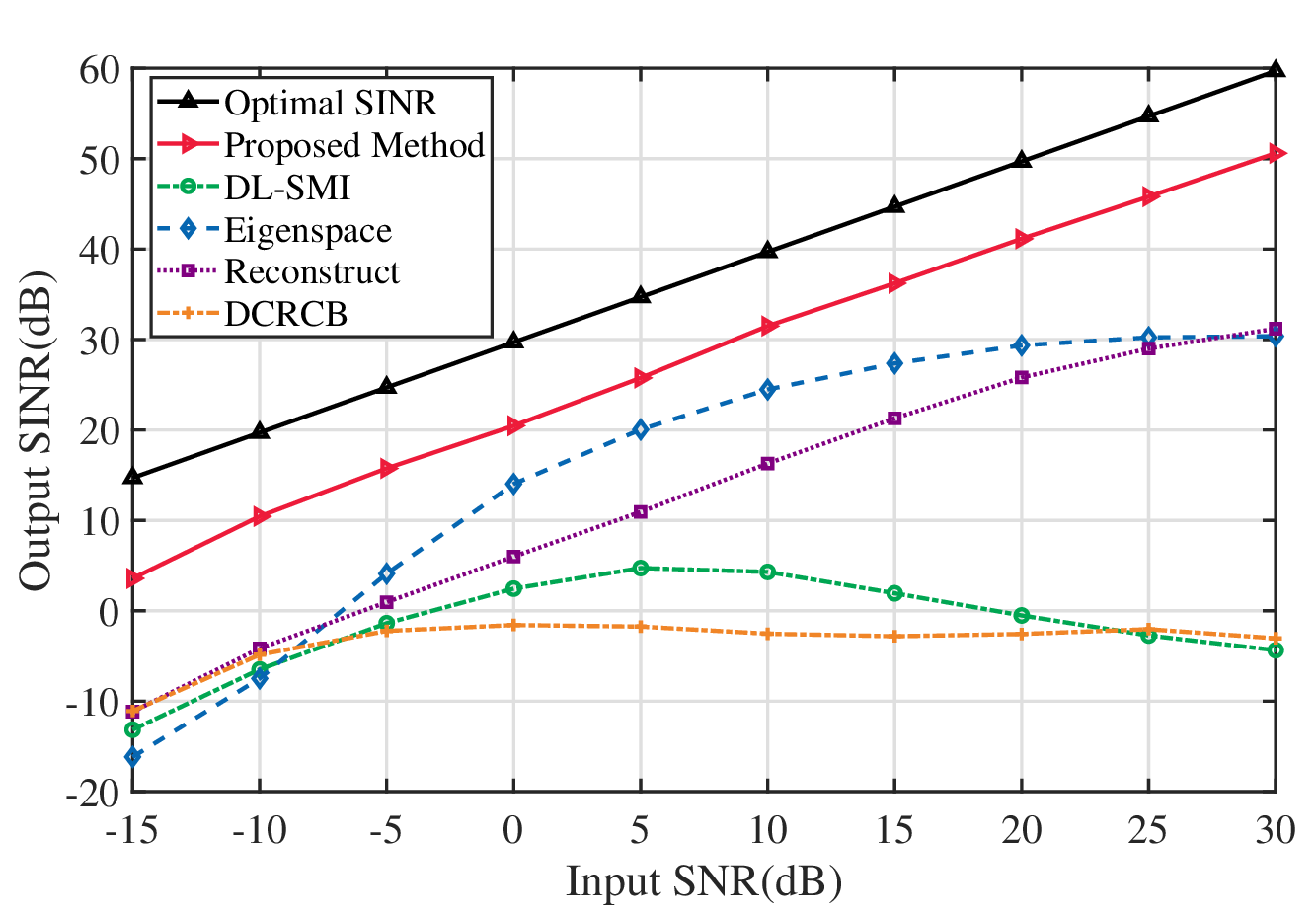}}
		\subfigure {\includegraphics[width=0.65\textwidth]{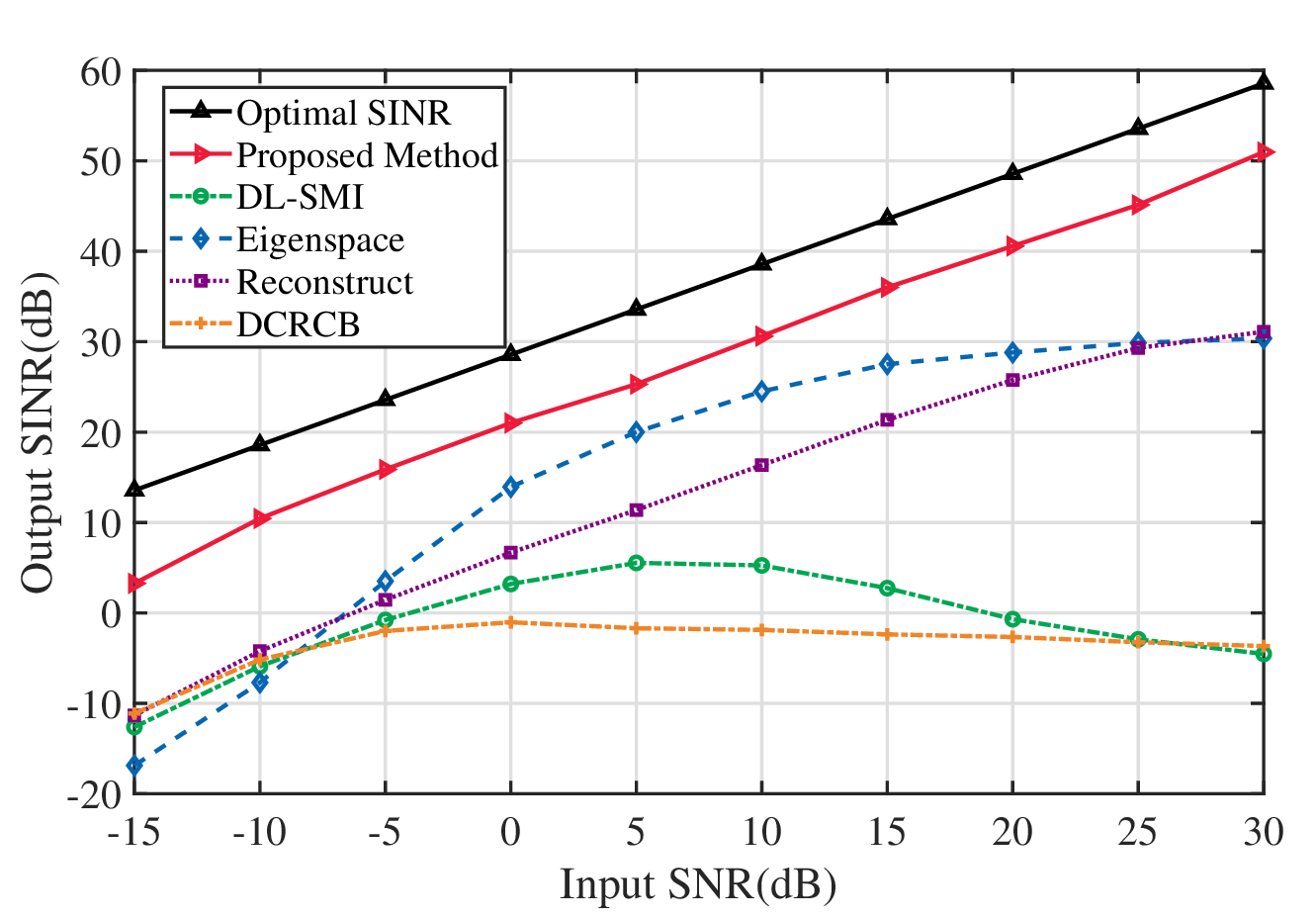}}
	\subfigure {\includegraphics[width=0.65\textwidth]{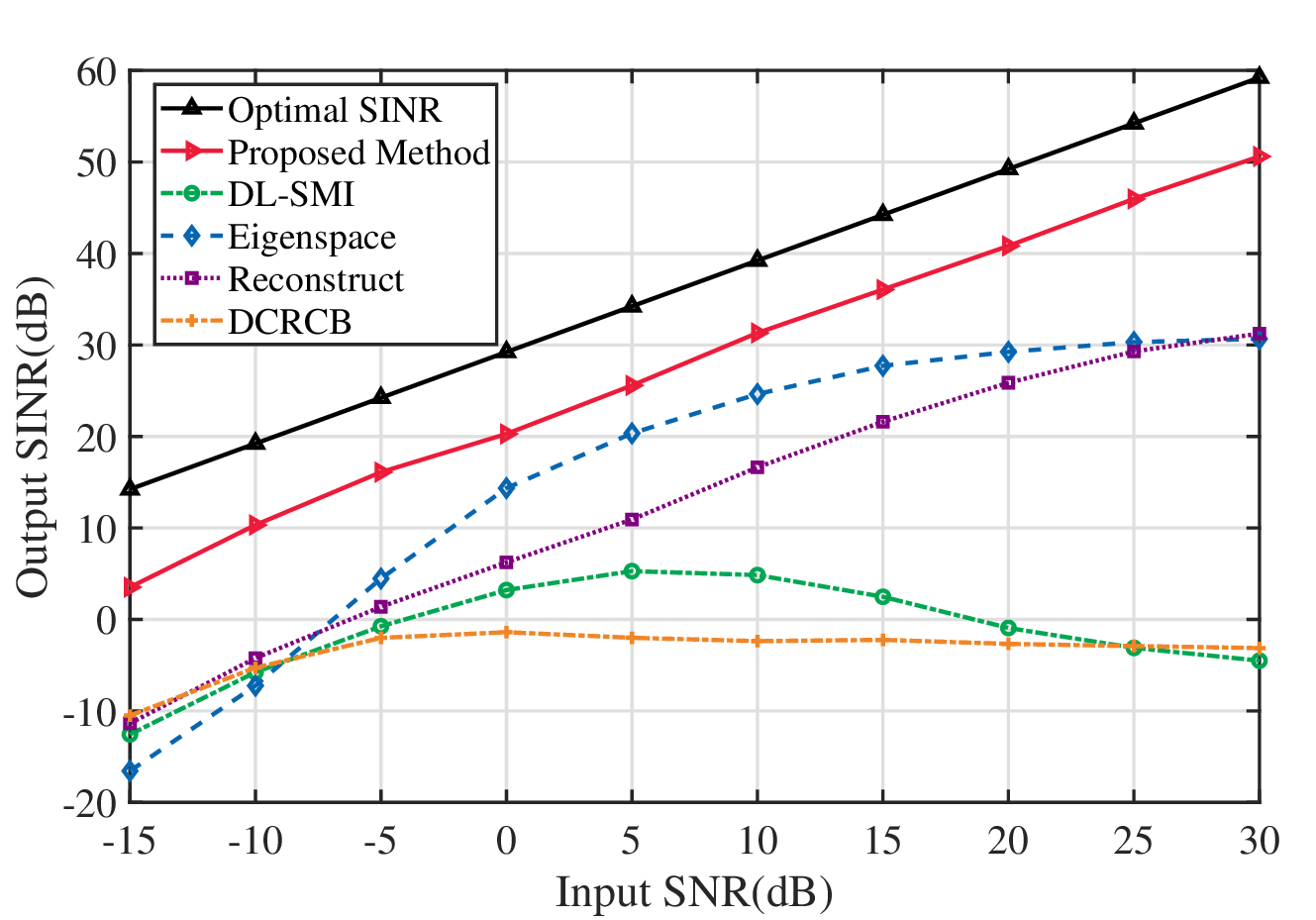}}
	\caption{ Output SINR versus the input SNR for third scenario, (a) $ 0.54\lambda $ element spacing, (b) $ 0.5\lambda $, (c) $ 0.45\lambda $.}
	\label{f.15}
\end{figure}

In the above scenarios, the antenna array is assumed to be ideal ($ \boldsymbol{C}=\boldsymbol{I} $). However, these assumptions are not fully satisfied in practice due to uncertainties in element position and violation of the isotropic hypothesis. In this example, $ \boldsymbol{C} $ is obtained by HFSS. The array configuration for the model in \cite{8318636} is shown in Fig.\ref{f.11}. To examine the effect of mutual coupling for FCA, three element spacing cases ($ 0.54\lambda ,0.5\lambda $ and $0.45\lambda $, corresponding to different $ R_1$ and $ R_2 $, and simulation of the expansion, initial, and shrinkage state of SWMA) are considered to be associated with ACP errors. The other parameters are the same with example 2.
Fig.\ref{f.15} illustrates the performance of competing beamformers with $ 0.54\lambda ,0.5\lambda $ and $0.45\lambda $ element spacing at input SNR=20 dB, respectively.

Some observations are made from the figures:
\begin{itemize}
	\item The performance of all beamformers is degraded considering mutual coupling effects at lower input SNR. 
	\item The proposed method maintains the output SINR after compensating the mutual coupling effect, which is only reduced by 2 dB compared with Fig.\ref{f.5}, while the other methods degrade by 10 dB without considering mutual coupling when input SNR is less than -10 dB. This is due to the radiation energy consumption in the adjacent elements.
	\item When the spacing between the elements is approximately half of the wavelength, minimal fluctuations in spacing have negligible impact on the resulting SINR, which indicates that variations in mutual coupling between the array elements are also minor. This fact also shows the feasibility of off-line measurement of MCM.
\end{itemize}


After compensating, our method ensures that the output SNR of the beamformer does not deteriorate with the mutual coupling.
Thus, the effectiveness of our proposed method is verified.

\section{CONCLUSION}
Considering the mutual coupling effect together with mixed mismatches in beamforming for FCA, we propose a robust beamforming algorithm based on multiple domain INCM reconstruction and alternative optimization. We verified from computer simulations that the proposed method outperforms other robust beamformers in respect of SINR output and beam focusing to the target. In the future, we will investigate a more complex situations to reconstruct the optimized object and figure out a simplified algorithm to calculate INCM. Moreover, the proposed algorithms show superior attitude against mutual coupling by introducing mutual coupling matrix into the beamformer while the performance of other robust approaches is unstable.
Future work may analyse other scenarios such as near field effect, multiple control parameter FCA, and a theoretical demonstrator.
\vspace*{-.5pc}


%
\section*{APPENDIX}
\subsection{Derivation of \eqref{27} and \eqref{28}}
\label{app1}
Considering the structure of the FCA illustrated in Fig.\ref{ring_array}, the SV of SOI with the mismatch of ACPs can be denoted as
\begin{equation}\label{ap1}
\bar{\mathbf{a}}\left( \theta _1,r_1,r_2 \right) =\bar{\mathbf{a}}_{pr}\left( \theta _1,\bar{r}_1,\bar{r}_2 \right) \odot \boldsymbol{\alpha }\left( \theta _1,x,y \right) 
\end{equation}
where $ \left| x \right|=\left| r_1-\bar{r}_1 \right| , \left| y \right|=\left| r_2-\bar{r}_2 \right|$. $ \bar{\mathbf{a}}_{pr}\left( \theta _1,\bar{r}_1,\bar{r}_2 \right)  $ is the presumed SV without mismatch of ACPs. $ \boldsymbol{\alpha }\left( \theta _1,x,y \right)  $ is the deviation between the real and the assumed SV.
\begin{equation}
\label{ap2}
\begin{aligned}
\bar{\mathbf{a}}_{pr}=&\left[ \begin{matrix}
e^{jk\boldsymbol{p}_1\left( \bar{r}_1,\bar{r}_2 \right) \cdot \boldsymbol{u}\left( \theta _i \right)}&		\cdots&		e^{jk\boldsymbol{p}_S\left( \bar{r}_1,\bar{r}_2 \right) \cdot \boldsymbol{u}\left( \theta _i \right)}\\
\end{matrix} \right] ^T\\
\boldsymbol{\alpha }=&\left[ \begin{matrix}
e^{jk\boldsymbol{p}_1\left( x,y \right) \cdot \boldsymbol{u}\left( \theta _i \right)}&		\cdots&		e^{jk\boldsymbol{p}_S\left( x,y \right) \cdot \boldsymbol{u}\left( \theta _i \right)}\\
\end{matrix} \right] ^T\\
\end{aligned}
\end{equation}
Then using (\ref{ap1}), (\ref{28}) can be reformulated as
\begin{equation}\label{ap3}
p_{2}^{\left( m \right)}: \left\{ \begin{array}{l}
\underset{x,y}{\min}\,\,\,\boldsymbol{\alpha }^H\hat{\boldsymbol{R}}_{eq}\boldsymbol{\alpha }+\mathrm{Re}\{ ( \mathbf{e}_{\bot}^{\left( m \right)} ) ^H\grave{\boldsymbol{R}}\mathrm{diag}\left( \bar{\mathbf{a}}_{pr}^{H} \right) \boldsymbol{\alpha } \}\\
s.t.   \boldsymbol{\alpha }^H\mathrm{diag}\left( \bar{\mathbf{a}}_{pr}^{H} \right) \mathbf{e}_{\bot}^{\left( m \right)}=0,\left| x \right|\leqslant l_1,\left| y \right|\leqslant l_2\\
\end{array} \right. 
\end{equation}
where we using the following notation
\begin{equation}\label{ap4}
\hat{\boldsymbol{R}}_{eq}=\mathrm{diag}\left( {\bar{\mathbf{a}}_{pr}}^H \right) \boldsymbol{C}^H\hat{\boldsymbol{R}}^{-1}\boldsymbol{C}\mathrm{diag}\left( \bar{\mathbf{a}}_{pr} \right) 
\end{equation}
By virtue of the \textsl{inner point method}, the penalty function $ f\left( x,y \right) $ can be represented as (\ref{ap5}). Then (\ref{ap3}) transformed into the unconstrained optimization problem $ \min  f\left( x,y \right)  $ and \textsl{Newton's method} can be applied in this case.
	\begin{equation}\label{ap5}
\begin{aligned}
f\left( x,y \right) =&\boldsymbol{\alpha }^H\hat{\boldsymbol{R}}_{eq}\boldsymbol{\alpha }+\mathrm{Re}\{(\mathbf{e}_{\bot}^{\left( m \right)})^H\grave{\boldsymbol{R}}\mathrm{diag}\left( \bar{\mathbf{a}}^H \right) \boldsymbol{\alpha }\}+\\
&\epsilon ^{\left( k \right)}\boldsymbol{\alpha }^H\mathrm{diag}\left( \bar{\mathbf{a}}^H \right) \mathbf{e}_{\bot}^{\left( m \right)}(\mathbf{e}_{\bot}^{\left( m \right)})^H\mathrm{diag}\left( \bar{\mathbf{a}} \right) \boldsymbol{\alpha }-\\
&\epsilon ^{\left( k \right)}\ln \left( l_{1}^{2}-x^2 \right) -\epsilon ^{\left( k \right)}\ln \left( l_{2}^{2}-y^2 \right) -\\
&\epsilon ^{\left( k \right)}\ln\mathrm{(}-\mathrm{Re}\{(\mathbf{e}_{\bot}^{\left( m \right)})^H\grave{\boldsymbol{R}}\mathrm{diag}\left( \bar{\mathbf{a}}^H \right) \boldsymbol{\alpha }\}-(\mathbf{e}_{\bot}^{\left( m \right)})^H\grave{\boldsymbol{R}}\mathbf{e}_{\bot}^{\left( m \right)})\\
\end{aligned}
	\end{equation}
Here $ \epsilon ^{\left( k \right)},k=0,1,\cdots$, is a parameter sequence of barrier weight, which satisfies $ \epsilon ^{\left( k \right)}\rightarrow 0 $. We write the first derivative of $ f\left( x,y \right) $ in (\ref{ap6}) and (\ref{ap7}).
	\begin{equation}\label{ap6}
\begin{aligned}
\frac{\partial f}{\partial x}=&\mathrm{Re}\{{\dot{\boldsymbol{\alpha}}_x}^H\hat{\boldsymbol{R}}_{eq}\boldsymbol{\alpha }\}+\mathrm{Re}\{{\mathbf{e}^H}_{\bot}\grave{\boldsymbol{R}}\mathrm{diag}\left( \bar{\mathbf{a}}^H \right) \dot{\boldsymbol{\alpha}}_x\}+\\
&\epsilon ^{\left( k \right)}\mathrm{Re}\{{\dot{\boldsymbol{\alpha}}_x}^H\mathrm{diag}\left( \bar{\mathbf{a}}^H \right) \mathbf{e}_{\bot}^{\left( m \right)}(\mathbf{e}_{\bot}^{\left( m \right)})^H\mathrm{diag}\left( \bar{\mathbf{a}} \right) \boldsymbol{\alpha }\}+\\
&\frac{2x}{l_{1}^{2}-x^2}\epsilon ^{\left( k \right)}-\frac{\mathrm{Re}\{(\mathbf{e}_{\bot}^{\left( m \right)})^H\grave{\boldsymbol{R}}\mathrm{diag}\left( \bar{\mathbf{a}}^H \right) \dot{\boldsymbol{\alpha}}_x\}}{\mathrm{Re}\{(\mathbf{e}_{\bot}^{\left( m \right)})^H\grave{\boldsymbol{R}}\mathrm{diag}\left( \bar{\mathbf{a}}^H \right) \boldsymbol{\alpha }\}+(\mathbf{e}_{\bot}^{\left( m \right)})^H\grave{\boldsymbol{R}}\mathbf{e}_{\bot}^{\left( m \right)}}\epsilon ^{\left( k \right)}\\
\end{aligned}
	\end{equation}
	\begin{equation}\label{ap7}
\begin{aligned}
\frac{\partial f}{\partial y}=&\mathrm{Re}\{{\dot{\boldsymbol{\alpha}}_x}^H\hat{\boldsymbol{R}}_{eq}\boldsymbol{\alpha }\}+\mathrm{Re}\{{\mathbf{e}^H}_{\bot}\grave{\boldsymbol{R}}\mathrm{diag}\left( \bar{\mathbf{a}}^H \right) \dot{\boldsymbol{\alpha}}_y\}+\\
&\epsilon ^{\left( k \right)}\mathrm{Re}\{{\dot{\boldsymbol{\alpha}}_y}^H\mathrm{diag}\left( \bar{\mathbf{a}}^H \right) \mathbf{e}_{\bot}^{\left( m \right)}(\mathbf{e}_{\bot}^{\left( m \right)})^H\mathrm{diag}\left( \bar{\mathbf{a}} \right) \boldsymbol{\alpha }\}+\\
&\frac{2y}{l_{1}^{2}-y^2}\epsilon ^{\left( k \right)}-\frac{\mathrm{Re}\{(\mathbf{e}_{\bot}^{\left( m \right)})^H\grave{\boldsymbol{R}}\mathrm{diag}\left( \bar{\mathbf{a}}^H \right) \dot{\boldsymbol{\alpha}}_y\}}{\mathrm{Re}\{(\mathbf{e}_{\bot}^{\left( m \right)})^H\grave{\boldsymbol{R}}\mathrm{diag}\left( \bar{\mathbf{a}}^H \right) \boldsymbol{\alpha }\}+(\mathbf{e}_{\bot}^{\left( m \right)})^H\grave{\boldsymbol{R}}\mathbf{e}_{\bot}^{\left( m \right)}}\epsilon ^{\left( k \right)}\\
\end{aligned}
	\end{equation}
where we use the notation of  
\begin{equation}\label{ap8}
\begin{aligned}
\dot{\boldsymbol{\alpha}}_x&=\frac{\partial \boldsymbol{\alpha }\left( x,y \right)}{\partial x}=jk\tilde{\boldsymbol{\mu}}_{\boldsymbol{u}}\odot \boldsymbol{\alpha }
\\
\dot{\boldsymbol{\alpha}}_y&=\frac{\partial \boldsymbol{\alpha }\left( x,y \right)}{\partial y}=jk\tilde{\boldsymbol{\nu}}_{\boldsymbol{u}}\odot \boldsymbol{\alpha }
\end{aligned}
\end{equation}
and $ \boldsymbol{u}=\left[ \begin{matrix}
\cos \theta&		\sin \theta\\
\end{matrix} \right]^T  $, 
\begin{equation}\label{ap9}
\tilde{\boldsymbol{\nu}}_{\boldsymbol{u}}=\left[ \begin{matrix}
\boldsymbol{\nu }_{1}^{T}\boldsymbol{u}&		\cdots&		\boldsymbol{\nu }_{S}^{T}\boldsymbol{u}\\
\end{matrix} \right] ^T,\tilde{\boldsymbol{\mu}}_{\boldsymbol{u}}=\left[ \begin{matrix}
\boldsymbol{\mu }_{1}^{T}\boldsymbol{u}&		\cdots&		\boldsymbol{\mu }_{S}^{T}\boldsymbol{u}\\
\end{matrix} \right] ^T
\end{equation}
where
\begin{equation}\label{ap10}
\begin{aligned}
\boldsymbol{\mu }_{i}^{T}=\left\{ \begin{array}{l}
\left[ \begin{matrix}
\sin \beta _i&		\cos \beta _i\\
\end{matrix} \right] ^T,i=-1,\cdots ,-M\\
\left[ \begin{matrix}
0&		1\\
\end{matrix} \right] ,i=1,\cdots ,N\\
\end{array} \right. 
\\
\boldsymbol{\nu }_{i}^{T}=\left\{ \begin{array}{l}
\left[ \begin{matrix}
0&		0\\
\end{matrix} \right] ,i=-1,\cdots ,-M\\
\left[ \begin{matrix}
\sin \beta _i&		\cos \beta _i-1\\
\end{matrix} \right] ^T,i=1,\cdots ,N\\
\end{array} \right. 
\end{aligned}
\end{equation}
According to the general form of Newton's method \cite{KuhnHaroldW}, the optimal estimated ACP can be obtained by
\begin{equation}\label{ap11}
\boldsymbol{p}^{k+1}=\boldsymbol{p}^k-\varrho  ^k\left( \nabla ^2f\left( x,y \right) \right) ^{-1}\nabla f\left( x,y \right) 
\end{equation}
where, $ \varrho  ^k $ is the $ k $-{th} step length coefficient of user choice, $ k $ denotes iteration times and
\begin{align}
\boldsymbol{p}^k=&\left[ \begin{matrix}
x^k&		y^k\\
\end{matrix} \right] ^T
\\
\label{ap12}
\nabla f\left( x,y \right) =&\left[ \begin{matrix}
\frac{\partial f\left( x,y \right)}{\partial x}&		\frac{\partial f\left( x,y \right)}{\partial y}\\
\end{matrix} \right] ^T
\\
\label{ap13}
\nabla ^2f\left( x,y \right) =&\left[ \begin{matrix}
\frac{\partial ^2f\left( x,y \right)}{\partial x^2}&		\frac{\partial ^2f\left( x,y \right)}{\partial x\partial y}\\
\frac{\partial ^2f\left( x,y \right)}{\partial y\partial x}&		\frac{\partial ^2f\left( x,y \right)}{\partial y^2}\\
\end{matrix} \right] 
\end{align}




  \bibliographystyle{elsarticle-num} 
    \bibliography{Robust}


%
%
%
\end{document}